\documentclass[aps,prd,floatfix,groupedaddress,nofootinbib,superscriptaddress,twocolumn,notitlepage]{revtex4-2}

\usepackage{amsmath,amssymb,bm,color,float,graphicx,mathrsfs}
\usepackage[hidelinks]{hyperref}
\hypersetup{colorlinks=true,linkcolor=blue,citecolor=blue,urlcolor=blue}
\usepackage{Macro}

\begin{document}

\title{Probing inflationary gravitational waves with cross-correlations: improved forecasting and validation with simulations}

\author{Toshiya Namikawa}
\affiliation{Center for Data-Driven Discovery, Kavli IPMU (WPI), UTIAS, The University of Tokyo, Kashiwa, 277-8583, Japan}
\affiliation{Department of Applied Mathematics and Theoretical Physics, University of Cambridge, Wilberforce Road, Cambridge CB3 0WA, United Kingdom}
\affiliation{Kavli Institute for Cosmology, University of Cambridge, Madingley Road, Cambridge CB3 0HA, United Kingdom}
\author{Irene Abril-Cabezas}
\author{Blake D. Sherwin}
\affiliation{Department of Applied Mathematics and Theoretical Physics, University of Cambridge, Wilberforce Road, Cambridge CB3 0WA, United Kingdom}
\affiliation{Kavli Institute for Cosmology, University of Cambridge, Madingley Road, Cambridge CB3 0HA, United Kingdom}

\date{\today}

\begin{abstract}
We present a follow-up study to the method recently proposed by Namikawa and Sherwin (2023) to probe gravitational waves using cross-correlations between two cosmic microwave background (CMB) $B$-modes and a large-scale structure tracer. We first improve on the previous forecast by including the impact of CMB component separation and find that, if the tensor-to-scalar ratio is $r=0$, we can achieve $\sigma_r\simeq3.6\times10^{-3}$ by combining upcoming experiments, i.e., LiteBIRD, CMB-S4 and the Advanced Simons Observatory. With a more futuristic experiment, we can achieve even tighter constraints on $r$ if improved delensing can be realized. Using a simulated analysis pipeline, we also explore possible biases from higher-order terms in the lensing potential, which were previously not examined in detail. We find that these bias terms are negligible compared to a detectable signal from inflationary gravitational waves. Our simulated results confirm that this new method is capable of obtaining powerful constraints on $r$. The method is immune to Gaussian Galactic foregrounds and has a different response to non-Gaussian Galactic foregrounds than the $B$-mode power spectrum, offering an independent cross-check of $r$ constraints from the standard power spectrum analysis. 
\end{abstract} 


\maketitle

\section{Introduction} \label{sec:intro}

Measurements of the cosmic microwave background (CMB) anisotropies have played a key role in developing our standard cosmological model. In upcoming decades, measuring the polarization of the CMB will be at the forefront of observational cosmology. 
In particular, detecting or constraining the odd-parity ($B$-mode) component of CMB polarization will be crucial, as it provides a unique probe of inflationary gravitational waves (IGWs) and offers new insights into the early Universe~\cite{Kamionkowski:1996:GW,Seljak:1996:GW}.

To date, no observations have confirmed the presence of IGWs; only upper limits on their amplitude have been established. The BICEP/Keck Array collaborations recently constrained the tensor-to-scalar ratio, $r$, at a pivot scale of $0.05\,$Mpc$^{-1}$, finding $r < 0.036$ ($2\,\sigma$) for a fixed cosmology \cite{BK13,BK15:data}. Meanwhile, Ref.~\cite{Tristram:2021:r}, which combines BICEP/Keck Array data with Planck Public Release 4 and baryon acoustic oscillations, reports a slightly tighter bound of $r < 0.032$ ($2\sigma$) by simultaneously constraining cosmological parameters.
Several ongoing and upcoming CMB experiments aim to detect IGW $B$-modes within the next decade, including the BICEP Array~\cite{BICEPArray}, Simons Array~\cite{SimonsArray}, Simons Observatory (SO) \cite{SimonsObservatory}, LiteBIRD \cite{LiteBIRD:2022:PTEP}, and CMB-S4 \cite{CMBS4}. 

High-precision measurements of large-scale $B$-modes can place tight constraints on $r$ \cite{CMBS4:r-forecast}. However, this precision is fundamentally limited by other sources of $B$-modes, primarily Galactic foregrounds~\cite{B2I}. At large scales, polarized Galactic emission dominates the signal and must be accurately mitigated. Although various methods have been developed to address this challenge~\cite{WMAP:2003:Bennett,Tegmark:2003:HILC,Delabrouille:2008qd,Stompor:2008:FGbuster,Katayama:2011eh,Ichiki:2018:delta-map}, the development of new methods for constraining IGWs with different or lower sensitivity to foregrounds is well motivated.

Reference~\cite{NS23} (hereafter NS23) recently proposed a novel method to measure IGWs using the gravitational lensing of IGW $B$-modes. This approach utilizes the fact that lensing distorts IGW $B$-modes, inducing correlations between the IGW $B$-modes at different angular scales. These correlations reflect the strength of the lensing effect along each line of sight and trace anisotropies in the CMB lensing potential. The resulting nonzero three-point correlation between two lensed IGW $B$-modes and an estimate of the CMB lensing potential, obtained from CMB lensing reconstruction and/or a correlated 
large-scale structure tracer, offers a new avenue for constraining $r$. Unlike traditional power spectrum analyses, this approach is expected to be largely immune to Galactic foregrounds, provided the lensing mass map remains almost uncontaminated by foregrounds. 
Even if foregrounds contribute significantly to our observable due to their non-Gaussianity, we expect that the foregrounds’ impact on our observable will differ significantly from their effect on the $B$-mode power spectrum, although an investigation of the impact of the foreground non-Gaussianity is necessary to test this expectation in detail.
Thus, the NS23 method has the potential to serve as a valuable cross-check for other constraints on $r$, with entirely different systematic errors. 

This paper is one of the follow-up papers which study in further detail the potential of the approach proposed in NS23. In this paper, we improve on NS23 in two ways: first, we update previous forecasts by including a noise spectrum that incorporates component separation; second, we perform a full simulated validation of the method, quantifying possible biases arising from higher-order terms in the lensing potential as discussed in NS23.

Throughout this paper, we compute the CMB power spectra with \textsc{camb} \citep[][]{CAMB}.

\section{Method}

\subsection{Lensed CMB polarization}

We begin with a brief introduction to the properties of lensed CMB polarization anisotropies. CMB observations obtain the Stokes $Q$ and $U$ parameters on the unit sphere at each line-of-sight direction $\hatn$. We then define the $E$- and $B$-modes of the CMB polarization from the Stokes parameters as \cite{Zaldarriaga:1996:EBdef}
\al{
    E_{\l m} \pm\iu B_{\l m} = - \Int{2}{\hatn}{} (Y_{\l m}^{\pm 2}(\hatn))^* [Q\pm\iu U](\hatn)
    \,, 
}
where $Y_{\l m}^2(\hatn)$ is the spin-2 spherical harmonics. 

CMB photons are influenced by the gravitational potential of the large-scale structure along their path. Lensing distortion of the CMB map is well approximated as a remapping of the primary CMB map with a deflection angle $\bm{d}(\hatn)$ (see Ref.~\cite{Lewis:2006fu} for a review). The lensing distortion then mixes the $E$- and $B$-modes at different angular scales \cite{Zaldarriaga:1998:LensB}. Introducing the lensing potential $\phi$, with $\bm{d}=\bn\phi$, the $B$-modes are modified as follows \cite{Hu:2000:cmblens,Namikawa:2011:curlrec}: 
\al{
	\tilde{E}_{\l m} &= E_{\l m} + \sum_{LM\l'm'}(-1)^m\Wjm{\l}{L}{\l'}{-m}{M}{m'}
	\notag \\
	&\qquad \times \grad_{LM} (E_{\l'm'}W^{+}_{\l L\l'} + \iu B_{\l'm'}W^{-}_{\l L\l'})
	\,, \label{Eq:E-x} \\ 
    \tilde{B}_{\l m} &= B_{\l m}+\sum_{LM\l'm'}(-1)^m\Wjm{\l}{L}{\l'}{-m}{M}{m'} 
    \notag \\
    &\qquad \times \grad_{LM}(-\iu E_{\l'm'}W^{-}_{\l L\l'}+B_{\l'm'}W^{+}_{\l L\l'}) 
    \,, \label{Eq:B-x}
}
where $\phi_{LM}$ are the spherical harmonic coefficients of the lensing potential, and we ignore the higher-order terms of $\grad$. We have defined
\al{
    W^{\pm}_{\l L\l'} &\equiv - q^{\pm}_{\l L\l'} \sqrt{\frac{L(L+1)(2\l+1)(2L+1)(2\l'+1)}{4\pi}}   
    \notag \\
	&\qquad\times \frac{1}{2}\bigg[ \sqrt{(\l'-2)(\l'+3)} \Wjm{\l}{L}{\l'}{2}{1}{-3} 
    \notag \\
	&\qquad\qquad + \sqrt{(\l'+2)(\l'-1)} \Wjm{\l}{L}{\l'}{-2}{1}{1} 
	\bigg]
	\,,
}
with $q^\pm_{\l L\l'}\equiv [1\pm (-1)^{\l+L+\l'}]/2$.

\subsection{Method of NS23}

Next, we review the new approach proposed in NS23 for detecting IGWs. Using a quadratic estimator, we reconstruct the lensing potential scaled by the tensor-to-scalar ratio, $\psi \equiv r\phi$, solely from lensed $B$-modes. We then cross-correlate $\psi$ with a CMB lensing map or a large-scale structure tracer, $x$, that is strongly correlated with CMB lensing. 
We also discuss the relation to the bispectrum estimator in Sec.~\ref{sec:bispec}. We comment on the large-scale $B$-mode reconstruction using small-scale CMB anisotropies and a lensing mass tracer in Appendix \ref{app:A}. 

For a full-sky idealistic case, the estimator to reconstruct $\psi$ from the lensed $B$-modes is given by [NS23]
\al{
	\hat{\psi}^*_{LM}
	= \frac{1}{2}A^{\psi\psi}_L\sum_{\l\l'mm'}\Wjm{\l}{\l'}{L}{m}{m'}{M}
	f^\psi_{\l L\l'}\frac{\hat{B}_{\l m}}{\hC^{BB}_\l}\frac{\hat{B}_{\l'm'}}{\hC^{BB}_{\l'}}
	\,, \label{Eq:estg} 
}
where $\hat{B}_{\l m}$ is an observed $B$-mode multipoles, $\hC^{BB}_{\l}$ is the best estimate of the observed, noisy $B$-mode power spectrum and we introduce
\al{
    f^\psi_{\l L\l'} 
    &\equiv W^{+}_{\l L\l'}C^{BB,r=1}_{\l'} + p_{\l L\l'}W^{+}_{\l'L\l}C^{BB,r=1}_\l 
    \,, \label{Eq:kernel:BB:x}
}
with $p_{\l L\l'}\equiv(-1)^{\l+L+\l'}$. Here, $C_\l^{BB,r=1}$ is the IGW $B$-mode power spectrum with $r=1$. $A^{\psi\psi}_L$ is the estimator normalization and is given in the full-sky idealistic case by
\al{
	(A_L^{\psi\psi})^{-1} = \frac{1}{2L+1}\sum_{\l\l'}\frac{(f^\psi_{\l L\l'})^2}{2\hC^{BB}_\l\hC^{BB}_{\l'}} 
	\,. \label{Eq:Norm:psi}
}
The cross-power spectrum between the reconstructed $\psi$ field and a lensing-mass tracer is proportional to $r$. Thus, we can constrain $r$ by measuring the amplitude of the cross-power spectrum, $C_L^{\psi x}$. The expected $1\,\sigma$ uncertainties of $r$ from the cross-correlations between $\psi$ and $x$ is [NS23]
\al{
    \sigma^{-2}_r \simeq \sum_L (2L+1)f_{\rm sky}\frac{\rho^2_LC_L^{\phi\phi}}{A_L^{\psi\psi}}
    \,, \label{Eq:sigmar}
}
where $f_{\rm sky}$ is the sky fraction of a CMB polarization observation, $\rho_L\equiv C_L^{\phi x}/\sqrt{C^{\phi\phi}_L\hC^{xx}_L}$ is the correlation coefficient between the true CMB lensing potential and the observed mass tracer, and $\hC_L^{xx}$ is the observed lensing-mass power spectrum. 

\subsection{Relation to the bispectrum estimator} \label{sec:bispec}

The above method is equivalent to measuring the cross bispectrum between two CMB $B$-modes and a lensing-mass tracer (i.e., a large-scale structure tracer that is correlated with lensing), $x$. \footnote{A similar approach was used in Ref.~\cite{Namikawa:2021:mode} to constrain cosmic birefringence, where they analyzed the (odd-parity) bispectrum $\ave{\tilde{B}\tilde{B}x}$ induced by parity violation} To see this, we here consider the bispectrum, $\ave{\tilde{B}\tilde{B}x}$. Ignoring the higher-order terms of $\phi$, using Eq.~\eqref{Eq:B-x}, the bispectrum is given by 
\al{
    \mC{B}^{\l\l'L}_{mm'M} 
    &\equiv \ave{\tilde{B}_{\l m}\tilde{B}_{\l'm'}x_{LM}} 
    \notag \\ 
    &= \sum_{L'M'\l''m''}(-1)^m\Wjm{\l'}{L'}{\l''}{-m'}{M'}{m''}
    \notag \\
    &\times\ave{B_{\l m} (\grad_{L'M'}B_{\l''m''}W^{+}_{\l' L'\l''})x_{LM}} + (\l\leftrightarrow\l')
    \notag \\ 
    &= \Wjm{\l'}{L}{\l}{-m'}{-M}{-m}C_{\l}^{BB}C_L^{\grad x}
    W^{+}_{\l' L\l} + (\l\leftrightarrow\l')
    \notag \\ 
    &= r\Wjm{\l}{\l'}{L}{m}{m'}{M}C_L^{\grad x}f^\psi_{\l L\l'}
    \,. 
}
Thus, measuring the amplitude of the bispectrum places a constraint on $r$. 

We can estimate $r$ using the same approach that is used to constrain primordial non-Gaussianity from the bispectrum \cite{Komatsu:2003:KSW,Babich:2004,Creminelli:2005,Babich:2005}. The optimal estimator for the amplitude of the bispectrum is given in an idealistic full-sky case by 
\footnote{The realization-dependent term, usually referred to as the linear term in the context of bispectrum estimation \cite{Creminelli:2005}, that contains $\ave{B_{\l m}B_{\l'm'}}x_{LM}+\ave{B_{\l m}x_{LM}}B_{\l'm'}+\ave{B_{\l'm'}x_{LM}}B_{\l m}$, becomes zero in an idealistic full-sky case with no monopole components. The inclusion of this term, however, would be important if the covariance of the $B$-modes is not diagonal.} 
\al{
    \hat{r} = N\sum_{\l\l'Lmm'M}\frac{\mC{B}^{\l\l'L}_{mm'M}|_{r=1}}{2\hC^{BB}_\l\hC^{BB}_{\l'}\hC_L^{xx}} \hat{B}_{\l m}\hat{B}_{\l'm'}\hat{x}_{LM}
    \,, \label{Eq:r-estimator}
}
where we include $1/2$ instead of $1/6$ in the denominator for optimality \cite{Smith:2006ud,Mangilli:2013sxa}. 
The estimator normalization in the full-sky idealistic case is given by
\al{
    \frac{1}{N} &= \sum_{\l\l'Lmm'M}\frac{(\mC{B}^{\l\l'L}_{mm'M}|_{r=1})^2}{2\hC^{BB}_\l\hC^{BB}_{\l'}\hC_L^{xx}}
    \notag \\
    &= \sum_{\l\l'L}\frac{(f^\psi_{\l L\l'}C_L^{\grad x})^2}{2\hC^{BB}_\l\hC^{BB}_{\l'}\hC_L^{xx}}
    \,. 
}
The expected constraint on $\hat{r}$ is then given by \cite{Hu:2000:cmblens}
\al{
    \sigma^{-2}_r \simeq \frac{f_{\rm sky}}{N} = f_{\rm sky}\sum_L \rho^2_L C_L^{\phi\phi} \sum_{\l\l'}\frac{(f^\psi_{\l L\l'})^2}{2\hC^{BB}_\l\hC^{BB}_{\l'}}
    \,. \label{Eq:sigmar:bispec}
}
We ignore the non-Gaussian variance of the bispectrum. The same Eq.~\eqref{Eq:sigmar:bispec} is also obtained by substituting Eq.~\eqref{Eq:Norm:psi} into Eq.~\eqref{Eq:sigmar}. 

We note that the full-sky idealistic bispectrum estimator, Eq.~\eqref{Eq:r-estimator}, can be rewritten as
\al{
    \hat{r} &= N\sum_{LM}C_L^{\grad x}\frac{\hat{x}_{LM}}{\hC_L^{xx}}\sum_{\l\l'mm'} \Wjm{\l}{\l'}{L}{m}{m'}{M}f^\psi_{\l L\l'}\frac{\hat{B}_{\l m}\hat{B}_{\l'm'}}{2\hC^{BB}_\l\hC^{BB}_{\l'}} 
    \notag \\
    &\equiv N\sum_L\frac{(2L+1)(C_L^{\grad x})^2}{A_L^{\psi\psi}\hC_L^{xx}}\frac{1}{C_L^{\phi x}}\frac{1}{2L+1}\sum_M\hat{x}_{LM}\hat{\psi}^*_{LM}
    \,. \label{Eq:r-estimator:rewrite}
}
The above equation means that, to estimate $r$ via the bispectrum, we first compute $\hat{\psi}_{LM}$, cross-correlate it with $\hat{x}_{LM}$, estimate $r$ at each $L$ with $C_L^{\psi x}$, and combine these $r$ estimates over $L$ optimally by multiplying by the inverse variance. The NS23 method utilizes the bispectrum of the two $B$-modes and a lensing map, but NS23 does not show whether the estimator proposed in NS23 is an optimal bispectrum estimator. The above shows explicitly that the optimal $BBx$ bispectrum measurement is equivalent to the method of NS23.

\subsection{Comparison with the $B$-mode power spectrum case}

We compare the uncertainty, $\sigma_r$, obtained from our method with that from the standard $B$-mode power spectrum approach. In the Fisher matrix formalism, $\sigma_r$ in the case of the $B$-mode power spectrum is given by
\al{
    \sigma^{-2}_r \simeq \sum_\l \frac{2\l+1}{2}f_{\rm sky}\left(\frac{C^{BB,r=1}_{\l}}{\hC^{BB}_\l}\right)^2
    \,. \label{Eq:sigmar:BB} 
} 
Both Eqs.~\eqref{Eq:sigmar:bispec} and \eqref{Eq:sigmar:BB} scale inversely with the square of the observed $B$-mode power spectrum. This indicates that the constraints on $r$ from both methods similarly improve as the noise in the observed $B$-modes decreases, and as the level of residual lensing $B$-modes after delensing is reduced. However, our method utilizes a different multipole range of the observed $B$-modes, and the multipoles that most effectively constrain $r$ differ between the two approaches.

\section{Improved Forecasts} \label{sec:opt}

\begin{table}[t]
\centering
\begin{tabular}{c|c|c}
    Frequency [GHz] & $\sigma_{\rm P}$ [$\mu$K-arcmin] & $\theta$ [arcmin] \\ \hline
    27 & 11 & 91 \\
    39 & 7.6 & 63 \\
    93 & 1.4 & 30 \\
    145 & 1.6 & 17 \\
    225 & 4.2 & 11 \\
    280 & 10 & 9
\end{tabular}
\caption{Specific experimental setup for ASO-SAT, which scales the SO-SAT noise level by $1/\sqrt{2}$. Here, $\sigma_{\rm P}$ is the white noise level in the polarization map and $\theta$ is the full width at half maximum of the beam size. The experimental configuration of the SO-SAT is given  \href{https://www.dropbox.com/s/3psaobka9fgcyod/souk_forecast_suggestion.pdf?dl=0}{here}.
}
\label{table:ASO-SAT}
\end{table}

\begin{table}[t]
\centering
\begin{tabular}{c|c|c|c}
    Frequency [GHz] & $\sigma_{\rm P}$ [$\mu$K-arcmin] & $\theta$ [arcmin] & \# detectors \\ \hline
    40 & 37.42 & 70.5 & 48 \\
    50 & 33.46 & 58.5 & 24 \\
    60 & 21.31 & 51.1 & 48 \\
    68 & 16.87 & 45.0 & 168 \\
    78 & 12.07 & 40.0 & 192 \\
    89 & 11.30 & 37.0 & 168 \\
    100 & 6.56 & 34.0 & 510 \\
    119 & 4.58 & 30.0 & 632 \\
    140 & 4.79 & 27.0 & 510 \\
    166 & 5.57 & 28.9 & 488 \\
    195 & 5.85 & 28.6 & 620 \\
    235 & 10.79 & 24.7 & 254 \\
    280 & 13.80 & 22.5 & 254 \\
    337 & 21.95 & 20.9 & 254 \\
    402 & 47.45 & 17.9 & 338
\end{tabular}
\caption{Same as Table \ref{table:ASO-SAT} but for LiteBIRD. We mimic the experimental specification of Ref.~\cite{LiteBIRD:2022:PTEP}.}
\label{table:LiteBIRD}
\end{table}

\begin{table}[t]
\centering
\begin{tabular}{c|c|c|c}
    Frequency [GHz] & $\sigma_{\rm P}$ [$\mu$K-arcmin] & $\theta$ [arcmin] & \# detectors \\ \hline
    30 & 3.53 & 72.8 & 260 \\
    40 & 4.46 & 72.8 & 470 \\
    85 & 0.88 & 25.5 & 17000 \\
    95 & 0.78 & 22.7 & 21000 \\
    145 & 1.23 & 25.5 & 18000 \\
    155 & 1.34 & 22.7 & 21000 \\
    220 & 3.48 & 13.0 & 34000 \\
    270 & 5.97 & 13.0 & 54000
\end{tabular}
\caption{Same as Table \ref{table:ASO-SAT} but for S4D-SAT \cite{S4UD}. The number of detectors are taken from Ref.~\cite{CMBS4:r-forecast}.}
\label{table:S4D-SAT}
\end{table}

\begin{table}[t]
\centering
\begin{tabular}{c|cccccc}
    Experiment & $f_{\rm sky}$ & $\l_{\rm min}$ & $\l_{\rm max}$ & $L_{\rm min}$ & $L_{\rm max}$ & $\phi$ \\ \hline
    (A)SO-SAT & 0.1 & 30 & 500 & 10 & 500 & S4W-LAT \\
    LiteBIRD & 0.7 & 2 & 500 & 2 & 500 & S4W-LAT \\
    S4D-SAT & 0.03 & 50 & 500 & 10 & 500 & S4D-LAT 
\end{tabular}
\caption{Other experimental specifications for the $\psi$ observation.}
\label{table:experiments:psi}
\end{table}

\begin{table}[t]
\centering
\begin{tabular}{c|cccc}
    Experiment & $\sigma_{\rm P}$ [$\mu$K-arcmin] & $\theta$ [arcmin] & $\l^{\rm E}_{\rm min}$ & $\l^{\rm B}_{\rm min}$ \\ \hline
    S4W-LAT & 1.41 & 1 & 100 & 500 \\
    S4D-LAT & 0.40 & 1 & 100 & 500 
\end{tabular}
\caption{Experimental specifications for the $\phi$ observations. S4W-LAT mimics the S4 Wide-Deep Survey from Chile \cite{CMBS4}. S4D-LAT mimics the Ultra-Deep Survey described in Ref.~\cite{S4UD}. $\l^X_{\rm min}$ is the minimum multipole used for the lensing reconstruction.}
\label{table:experiments:phi}
\end{table}

In this section, we refine the forecast results of NS23 by incorporating component separation into the calculation. We also explore how much the results can be improved by optimizing the instrument to constrain $r$. We apply the harmonic internal-linear-combination (HILC) method (e.g., Refs.~\cite{Bennett:2003, Eriksen:2004, Remazeilles:2011, McCarthy:2024}) for component separation, assuming no bias to $C_L^{\psi x}$ from residual foregrounds, as these residuals are not expected to contribute to the correlations.

\begin{table}[t]
    \centering
    \begin{tabular}{c|c|c|c}
        $\psi$ experiments & $\sigma_r$ & $A^{\l=100}_{\rm lens}$ & $\sigma_r^{A_{\rm lens}=1}$ \\ \hline
        SO-SAT + ASO-SAT & 0.012 & 0.21 & 0.045 \\ \hline
        LiteBIRD & 0.0077 & 0.21 & 0.018 \\ \hline
        S4D-SAT  & 0.0064 & 0.053 & 0.086 \\ 
                 & (0.0041) & (0.015) & \\ \hline
        Combined & 0.0047 & $-$ & \\
                 & (0.0035) & $-$ &
    \end{tabular}
    \caption{$1\,\sigma$ expected constraints on $r$. For the combined case, we use $f_{\rm sky}=0.5$ for LiteBIRD to avoid overlap between LiteBIRD and SO+ASO. The parentheses denote the values of $\sigma_r$ when we optimize the survey region and noise level of S4D. $A^{\l=100}_{\rm lens}$ is the value of the residual fraction of lensing $B$-mode power spectrum left over after delensing at $\l=100$. For a reference, we also show $\sigma_{r}$ without delensing in the last row.}
    \label{table:sigmar}
\end{table}

\subsection{Expected constraints from upcoming experiments}

To measure $\psi$, we need to observe $B$-modes at large angular scales, which will be achievable with LiteBIRD and CMB-S4. Additionally, we consider incorporating data from SO-like and Advanced Simons Observatory (ASO) like experiments, as SO-SAT and ASO-SAT data will be available alongside LiteBIRD and CMB-S4 (although we do not expect large improvements beyond CMB-S4). 

This section considers three experimental setups for measuring $\psi$:
\bi
\item SO+ASO,\footnote{In this paper, (A)SO denotes an A(SO)-like experiment whose experimental specifications are defined in Tables \ref{table:ASO-SAT} and \ref{table:experiments:psi}.}
\item LiteBIRD, and
\item the S4 Ultra-Deep Survey SAT (S4D-SAT).
\ei 
The frequency coverage, noise level, and beam size for the above experiments are summarized in Tables \ref{table:ASO-SAT}, \ref{table:LiteBIRD}, and \ref{table:S4D-SAT}, respectively. Additional experimental specifications are provided in Table \ref{table:experiments:psi}. For SO+ASO, we assume that SO and ASO observe the same 10\% sky region and the combined $\sigma_{\rm P}$ is $\sqrt{2/3}$ of those shown in Table \ref{table:ASO-SAT}.

Our forecast includes the component separation process for Galactic foregrounds in the $\psi$ reconstruction. We model the dust spectral energy distribution (SED) as a modified black-body spectrum and the synchrotron SED as a power-law spectrum (see, e.g., Ref.~\cite{Planck_2020_diffuse}). The power spectra of these foregrounds are modeled as power laws with amplitudes $A_d$ and $A_s$ and spectral tilts $\alpha_d$ and $\alpha_s$ for dust and synchrotron, respectively.
\bi
\item For (A)SO, we use the best-fit values from Ref.~\cite{Azzoni_2021}, based on realistic simulations within the SO footprint, which covers 10\% of the sky.
\item For LiteBIRD, we adopt the best-fit dust power spectrum parameters from \emph{Planck}
\footnote{\emph{Planck} (\url{https://www.esa.int/Planck}) is a European Space Agency (ESA) science mission with contributions from ESA member states, NASA, and Canada. The mask used in this work was accessed from the Planck Legacy Archive at \url{https://pla.esac.esa.int}.}
in regions where the 50\% \emph{Planck} Galactic mask is applied \cite{Planc_2020_dust}. For synchrotron, we follow Ref.~\cite{Fuskeland_2023} and use the Python Sky Model \cite{Thorne_2017} ``s1" model to obtain synchrotron power spectra within the same region.
\item For CMB-S4, we use the same foreground power spectra as SO but scaled by a factor of $0.1$ to approximate the foreground levels observed in the BICEP patch.
\ei 
Our results remain robust even if the foreground power spectrum is increased by a factor of $2$, indicating that the analysis is not highly sensitive to the foreground level.
We use the noise components of the $B$-mode power spectrum after the component separation in $B$-modes by adopting the HILC method:
\al{
    N^{BB,{\rm HILC}}_\l = \frac{\bm{w}^t_{\l}\bR{Cov}^N_\l\bm{w}_\l}{(\bm{1}^t\bm{w}_{\l})^2}
    \,,
}
where $\bR{Cov}^N_\l$ is the instrumental noise covariance matrix and $\bm{w}_\l$ is the optimal weight for the HILC:
\al{
    \bm{w}_\l = \bR{Cov}^{-1}_{\l}\bm{1} 
    \,.
}
Here, $\bm{1}=(1,1,\cdots,1)^T$ and the covariance matrix, $\bR{Cov}_\l$, contains CMB signals from lensing, $C_\l^{BB}$, instrumental noise, $N_{\l,\nu}^{BB}$, and Galactic foreground spectra, $C_{\l,\nu\mu}^{\rm FG}$. The instrumental noise is computed as a white noise with a Gaussian beam deconvolution. The covariance matrix of the $B$-modes is then given by
\al{
    \{\bR{Cov}_\l\}_{\nu\mu} = (C_{\l}^{BB} + N_{\l,\nu}^{BB})\delta_{\nu\mu} + C_{\l,\nu\mu}^{\rm FG}
    \,. 
}
We use the noise spectra after the HILC process, $N^{BB,{\rm HILC}}$, for computing the reconstruction noise of $\psi$. 
The HILC method is generally not sufficient for IGW constraints from $B$-mode power spectrum, but acceptable for methods that should be less susceptible to foregrounds.

After component separation, we assume that lensing $B$-modes are further removed from the large-scale observed $B$-modes for the $\psi$ measurement using the template delensing method. In this approach, lensing $B$-modes are reconstructed as a convolution of observed lensed $E$-modes and the lensing potential. The $E$-modes are estimated by combining both large- and small-scale experiments, while the lensing potential is measured by the small-scale experiment. For measuring $\phi$, we consider S4W-LAT for SO+ASO and LiteBIRD, and S4D-LAT for S4D-SAT. The experimental setup for these high-resolution observations, S4W-LAT and S4D-LAT, is summarized in Table \ref{table:experiments:phi}. The residual lensing $B$-mode power spectrum is computed analytically following Ref.~\cite{Smith:2010gu}. The delensing efficiency depends on the minimum multipole of the available $E$-modes. We assume that the $E$-modes are signal-dominated and combine data from both the small-scale experiment and LiteBIRD for $2 \leq \ell$, where LiteBIRD noise is modeled as $2.6\,\mu$K-arcmin white noise with a $30$-arcmin Gaussian beam. Additionally, for the small-scale experiment, we remove $E$-modes at $\ell < 100$ and $B$-modes at $\ell < 500$ since large-scale polarization measurements are challenging for high-resolution telescopes due to atmospheric $1/f$ noise.

The observed $B$-mode power spectrum in Eq.~\eqref{Eq:Norm:psi} is taken as the sum of the residual lensing $B$-mode power spectrum and the HILC noise power spectrum. The expected constraint on $r$ is then computed using Eq.~\eqref{Eq:sigmar}. The reconstructed $\phi$ is used for $x$. Table \ref{table:sigmar} summarizes the expected $1\,\sigma$ constraints on $r$, $\sigma_r$, assuming a fiducial value of $r=0$ and shows the level of residual lensing $B$-mode power spectrum after delensing at $\l=100$. We also summarize the case without delensing to see the importance of delensing for individual cases, although this case is not realistic, especially for the ground-based experiments with a small observing patch that is optimized for delensing. 
We also consider the scenario where all datasets are combined. Assuming that the sky regions of SO+ASO, LiteBIRD, and S4D-SAT do not overlap, we rescale $\sigma_r$ from LiteBIRD by $\sqrt{0.7/0.6}$, leading to a combined constraint of $\sigma_r = 0.0047$. 

\subsection{Sky coverage vs noise level}

\begin{figure}[t]
 \centering
 \includegraphics[width=86mm]{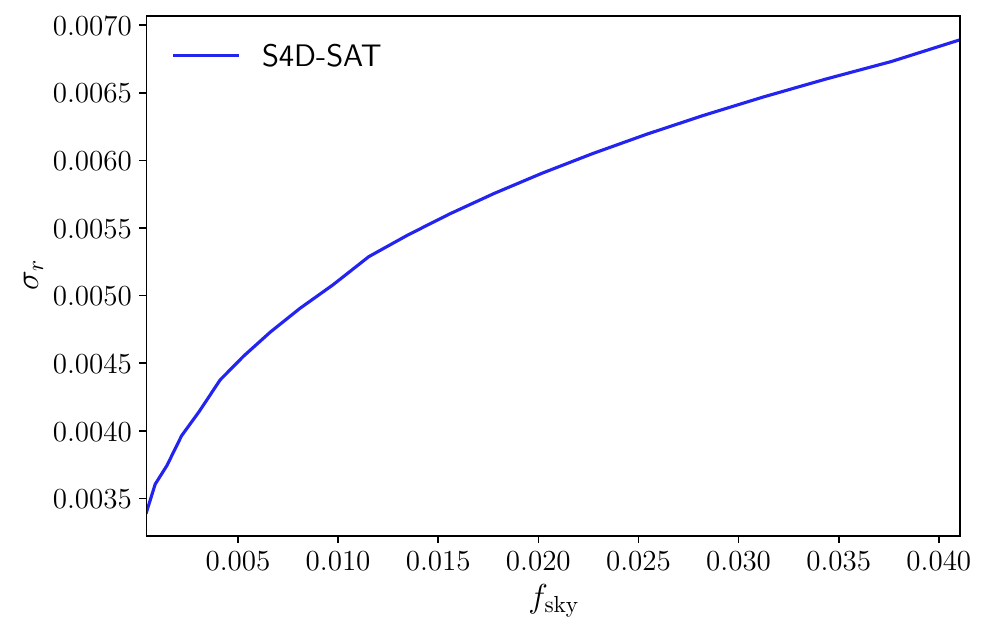}
 \caption{$1\sigma$ constraint on $r$ from the S4D-SAT case as a function of $f_{\rm sky}$, where we scale the noise level and sky coverage as $\sigma_{\rm P,\nu}=s\sigma_{\rm P,\nu}^{\rm fid}$ and $f_{\rm sky}=0.03\times s^2$, assuming a fixed observation time.}
 \label{fig:const:fsky}
\end{figure}

The sky coverage and noise level can be adjusted for a fixed observing time. Here, we simply assume that if the noise level is scaled by a factor $s$, the observing time per pixel changes uniformly by a factor of $s^2$. Consequently, the sky coverage can be increased or decreased accordingly, also scaling as $s^2$, due to the change in observation time per pixel. We compute $\sigma_r$ as a function of $s$ for the S4D-SAT case, scaling the noise level of both S4D-SAT and S4D-LAT by $s$ and the overlapping sky region by $s^2$. The foreground noise spectra remain unchanged across all-sky coverage cases, as their impact on the final $\sigma_r$ is negligible.

Figure~\ref{fig:const:fsky} presents the constraint on $r$ as a function of $f_{\rm sky}$. Tighter constraints can be achieved by observing a smaller region of the sky. We optimize this trade-off until $f_{\rm sky} \simeq 0.003$ (corresponding to $\sim 100\,$deg$^2$), where we obtain $\sigma_r = 0.0041$. Although further improvements in $\sigma_r$ are possible by reducing the observed sky area even more, achieving a delensing efficiency of $A_{\rm lens} \lesssim 0.01$ in such a configuration remains uncertain. Therefore, we do not explore this scenario further. When combining all experiments, we achieve a final constraint of $\sigma_r=0.0035$.

\subsection{Frequency bands}

Since our method is potentially less sensitive to foreground bias, we can operate with fewer observing frequencies than the standard analysis. Here, we investigate how much we can improve $\sigma_r$ by redistributing the number of detectors across frequency channels. In practice, reallocating detectors between frequency bands is not straightforward and the trade-offs are not strictly one-to-one. However, for simplicity, we assume that detectors can be freely reassigned between frequency bands. Keeping the total number of detectors fixed (a constraint adopted for simplicity rather than accuracy), and assuming that we adopt the HILC for the component separation, we determine the optimal detector distribution by minimizing the sum of the residual foreground $B$-mode power spectrum over multipoles, $\sum_{\ell}N_\ell^{BB,{\rm HILC}}$ where we recompute the HILC weight for each case. We assume no bias to $\psi\phi$ from residual foregrounds during the optimization.

After the optimization, we find that we need only $40\,$GHz, $140\,$GHz, $337\,$GHz, and $402\,$GHz frequency bands for LiteBIRD, and $30\,$GHz, $85\,$GHz, and $270\,$GHz frequency bands for S4D. Under this optimization, the noise power spectrum is reduced to approximately half of that in the fiducial detector distribution for LiteBIRD, yielding $\sigma_r = 0.0056$. For the S4D case, we obtain $\sigma_r = 0.0054$, which can be further improved to $\sigma_r = 0.0039$ by optimizing sky coverage. When combining all experiments, we achieve a final constraint of $\sigma_r = 0.0032$.

\subsection{Dependence on CMB multipole range}

\begin{figure}[t]
 \centering
 \includegraphics[width=86mm]{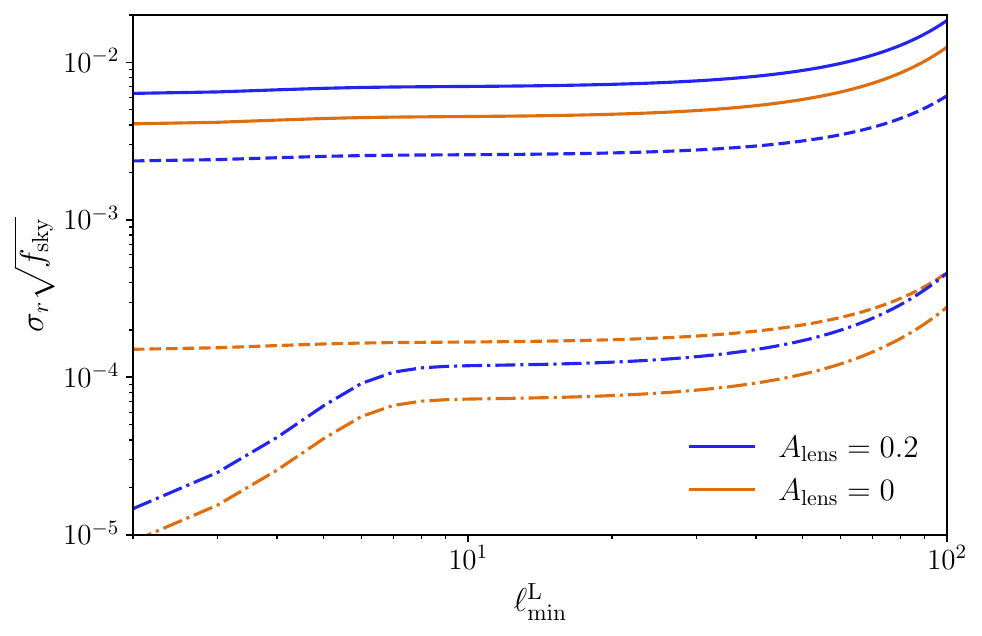}
 \caption{$1\sigma$ constraint on $r$ as a function of the minimum multipole of $B$-modes, $\l^{\rm L}_{\rm min}$, obtained from the LiteBIRD-like low-$\l$ experiment. We show two cases for the polarization noise level of the low-$\l$ experiment, $2.6\,\mu$K-arcmin (solid) and $0.5\,\mu$K-arcmin (dashed). The tracer used for cross-correlation is assumed to be an idealistic $\phi$ without the reconstruction noise. We set the lowest multipole of the lensing map to $\l^{\rm L}_{\rm min}$. We show two cases of delensing efficiency; $A_{\rm lens}=0.2$ and $0.0$. The $B$-mode multipoles at $\l\alt 30$ do not contribute to constraining $r$. For comparison, the dot-dashed lines show the constraints from the standard $B$-mode power spectrum approach with the same setup as that adopted in the solid line case.}
 \label{fig:const-r:lmin}
\end{figure}

\begin{figure}[t]
 \centering
 \includegraphics[width=86mm]{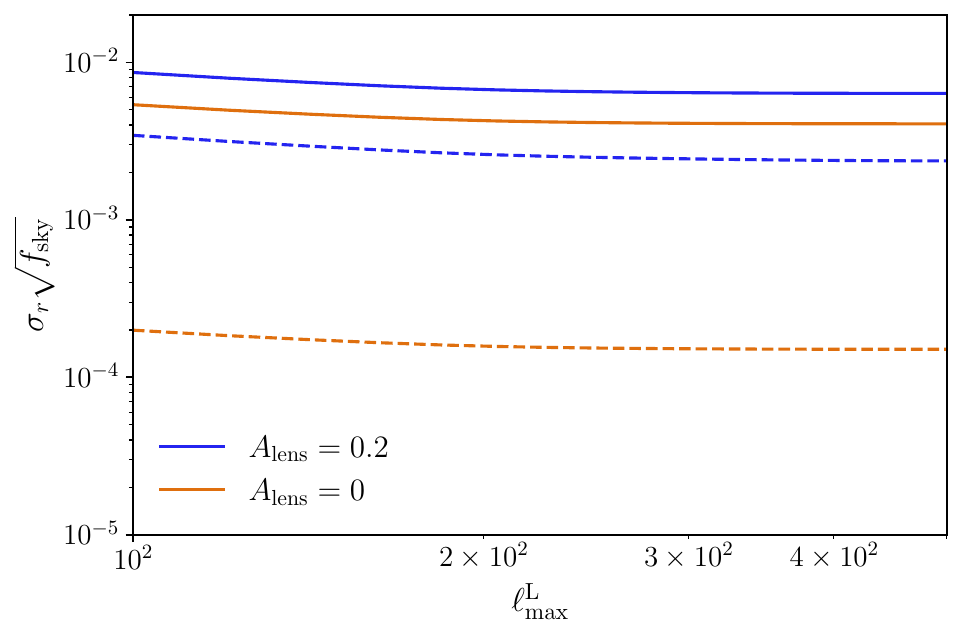}
 \caption{As in Fig.~\ref{fig:const-r:lmin}, but now varying the maximum multipole of $B$-modes, $\l^{\rm L}_{\rm max}$, for the $\psi$-reconstruction. The $B$-mode multipoles at $\l\agt 200$ do not contribute to constraining $r$. The constraint from the $B$-mode power spectrum does not change by varying $\l_{\rm max}^{\rm L}$ and is not shown.}
 \label{fig:const-r:lmax}
\end{figure}

The constraints on $r$ also depend on the multipole range of the CMB anisotropies and the reconstructed $\psi$ and $\phi$ fields. Figure~\ref{fig:const-r:lmin} shows $\sigma_r$ as a function of the minimum multipole for $B$-modes in the $\psi$ reconstruction, $\ell_{\rm min}^{\rm L}$, while keeping the maximum multipole fixed at $500$. We present two cases for the noise level of large-scale $B$-modes: $2.6\,\mu$K-arcmin (solid line) and $0.5\,\mu$K-arcmin (dashed line), assuming a Gaussian beam with a FWHM of $30$ arcmin. The correlation coefficients between the large-scale structure (LSS) tracer and the CMB lensing potential are set to unity, and we set the lowest multipole of the LSS tracer to $\l_{\rm min}^{\rm L}$. We assume that low-$\l$ $B$-modes are delensed using a lensing map and consider two cases for the residual lensing $B$-modes; $A_{\rm lens}=0.2$ and $A_{\rm lens}=0$. The results indicate that $B$-mode multipoles at $\l\lesssim 30$ do not contribute significantly to constraining $r$. Additionally, we verify that multipoles of the LSS tracer below $\ell \sim 30$ have a negligible impact on the constraint. This suggests that even a partial-sky observation from a ground-based experiment could potentially probe IGWs using our method.
For comparison, the dot-dashed lines show the constraints from the standard $B$-mode power spectrum approach with the same setup as that adopted in the solid line case. The large-scale multipoles, $\l\leq 10$, are important to constrain $r$ for the $B$-mode power spectrum approach.

Figure~\ref{fig:const-r:lmax} presents a similar analysis but now varies the maximum multipole of the large-scale $B$-modes for $\psi$ reconstruction, $\ell_{\rm max}^{\rm L}$, while fixing the minimum multipole at $\ell = 2$. We find that multipoles up to $\ell \sim 200$ are sufficient for near-optimal constraints on $r$ from $\psi$ reconstruction.

\subsection{Cosmic variance from nonzero $r$}

\begin{figure}[t]
    \centering
    \includegraphics[width=80mm]{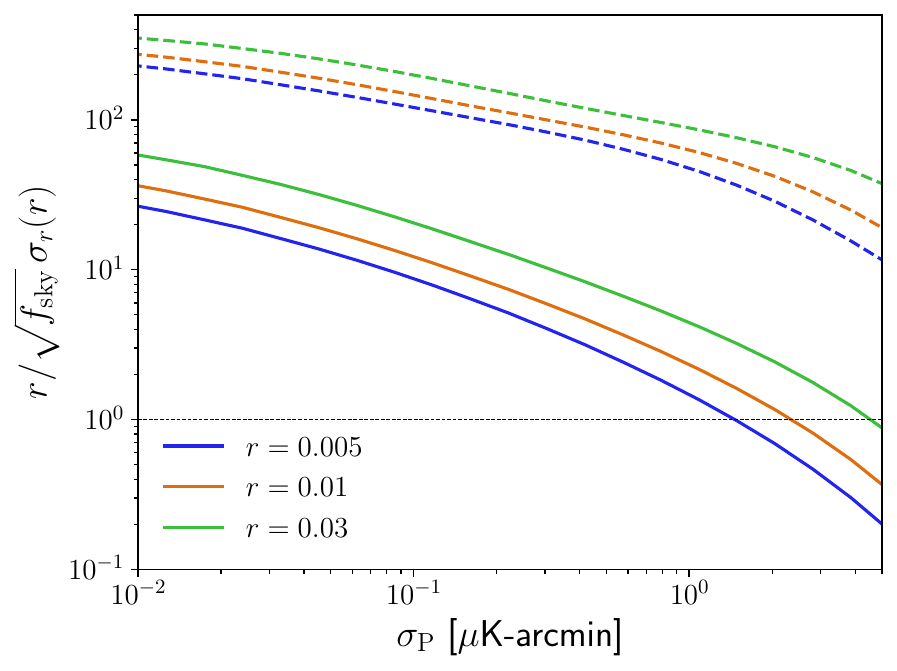}
    \caption{\textcolor{black}{Forecast for the SNR of the IGW signal in full-sky as a function of the post-component-separated white noise level in polarization, $\sigma_{\rm P}$, with our method (solid) and the standard analysis (dashed). We include the cosmic variance from the nonzero $r$. We assume an idealistic experiment that measures the polarization between $l=2$ and $4096$. The lensing map is reconstructed internally between $L=2$ and $L=4096$. We assume $\theta=1$ arcmin.}}
    \label{fig:nonzeror}
\end{figure}

So far, we have neglected cosmic variance from tensor perturbations, assuming $r=0$ in our error computation to assess the significance of rejecting the null hypothesis. Here, we compute $\sigma_r$ with the cosmic variance from tensor perturbations, assuming a nonzero true value of $r$. 

Figure~\ref{fig:nonzeror} shows the expected SNR of the IGW signal for a given $r$ as a function of the polarization noise level after component separation, assuming a white noise power spectrum. We compare results from our method and the standard power spectrum analysis. For the standard approach, we use Eq.~\eqref{Eq:sigmar:BB}. In both cases, the constraints are affected by the cosmic variance of $r$, with a more pronounced impact on the $B$-mode power spectrum than on $C_L^{\psi\phi}$. This is because the reionization bump in the $B$-mode power spectrum is particularly sensitive to the cosmic variance of $r$. 
Note that, unlike the forecasts presented in the previous subsections, the calculations here use the full multipole range for simplicity. In actual ground-based observations, however, atmospheric $1/f$ noise can limit the accessible minimum multipole. This poses a challenge for the standard approach, which relies on measuring the reionization bump---an additional source of information on $r$---at very low multipoles. In contrast, our method remains only marginally affected by this limitation, as the sensitivity to $r$ is preserved as long as the minimum multipole is $\ell_{\rm min} \lesssim 30$, as shown in the previous subsection.

\section{Reconstruction bias from higher-order lensing potential} 

\subsection{Higher-order biases from lensing}

A possible concern for our method is whether higher-order terms in $\phi$, especially ones that enter via the delensing procedure, might bias our measurements. Schematically, we may write modified versions of Eqs.~\eqref{Eq:E-x} and \eqref{Eq:B-x} as 
\al{
    \tilde E &= E + E\phi +B\phi+(n^E+E\phi^2+B\phi^2+E\phi^3+\cdots)
    \,, \\ 
    \tilde B &= B + B\phi +E\phi+(n^B+E\phi^2+B\phi^2+E\phi^3+\cdots)
    \,,
}
where the term in brackets now includes noise and higher-order terms. Note that it may be necessary to include higher orders in $\phi$ than typical lensing analyses since we are seeking to constrain very small values for $r\sim 10^{-3}$ (an additional small parameter in the problem).

We can construct a lensing $B$-mode template $B^t$ by applying the leading-order lensing operation (indicated with $\star$) to the observed Wiener-filtered lensed $E$-mode map; now working to higher order in $\phi$, we obtain
\al{
    B^t \equiv \tilde E\star \phi = E\phi +n^E \phi +E\phi^2 + B\phi^2+(E\phi^3+\cdots) 
    \,.
}
We may then delens by subtracting $\tilde B-B^t=B+B\phi+(E\phi-E\phi) +\cdots$; as indicated previously, if the LSS tracer $\phi$ that is used to construct the delensing template is perfectly correlated with the true lensing, then all leading order $E$-mode lensing signal is removed from the $B$-mode map. 

Only three higher-order terms may not be negligible when constructing the quadratic estimator $\hat{\psi}^*_{LM} \sim B_{\l m}B_{\l'm'}$ and correlating it with a LSS tracer $\phi$. 
\e
\item 
A term $\sim C_l^{EE}A^{1/2}_{\rm lens}\phi^4$, which arises from the residual lensing that has not been removed by delensing (due to an imperfect LSS tracer), correlating with the lensed $E$-mode term at $\mC{O}(\phi^2)$ and the LSS tracer; 
\item 
a term $\sim N^E\phi^3$, arising from the effect of $E$-mode noise in producing the delensing template; 
\item 
a term $\sim C_l^{EE}A_{\rm lens}\phi^3$ again arising from residual lensing. 
\ee 

The second and third terms resemble the so-called $N^{3/2}$ biases arising in lensing power spectrum measurements \cite{Boehm:2016,Boehm:2018}. The second, $E$-mode noise term is suppressed by the Wiener filter and is easily nulled by using different data splits with independent noise in constructing the delensing templates (if necessary, a very large number of splits can be used to recover all the signal-to-noise). 
The first and third terms are expected to be small, at least when delensing is efficient and $A_{\rm lens}$ is small. (We also note that the lensing bispectrum is much harder to detect than the lensing power spectrum \cite{Namikawa:2016:cmblbisp,Pratten:2016,Kalaja:2022}; indeed, it has not yet been detected.) We will test this expectation in the following section. Even if these terms are not entirely negligible, it would be straightforward to model and subtract any bias they produce.

\subsection{Testing our method with simulations}

\begin{figure*}[t]
\centering
\includegraphics[width=88mm]{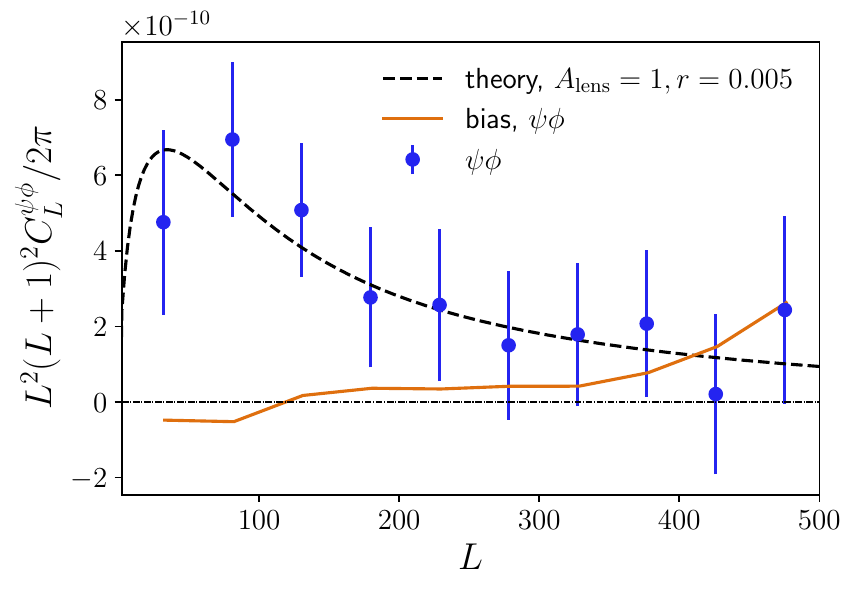}
\includegraphics[width=88mm]{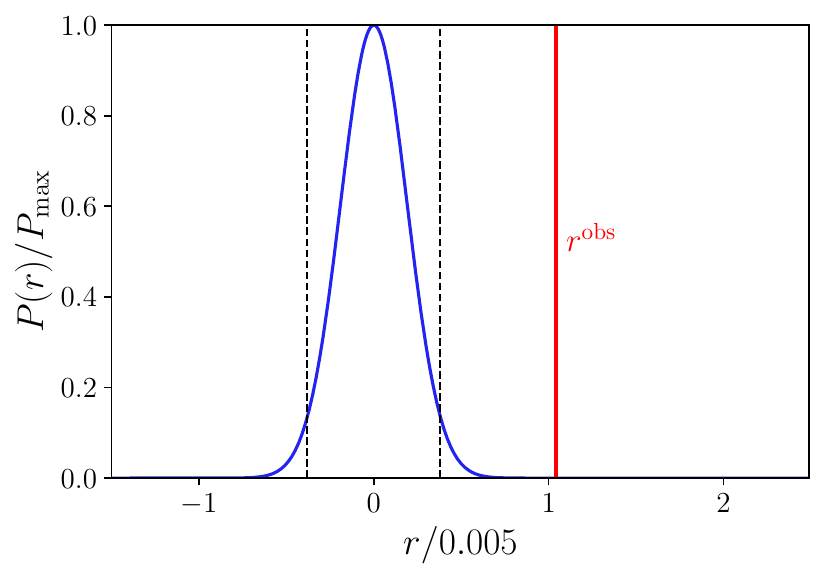}
\caption{
{\it Left}: cross-angular power spectrum between the reconstructed $\psi$ and input lensing potential, $C_L^{\psi\phi}$, obtained from an ideal full-sky simulation with $r=0.005$ (blue points), compared with the theoretical expectation (black dashed). The orange line denotes the bias arising from the higher-order lensing potential that is ignored in constructing the $\psi$ estimator. The bias and error bars are estimated from the $r=0$ simulation. We add a Gaussian white noise of $0.5\,\mu$K-arcmin with a $1$ arcmin Gaussian beam. 
{\it Right}: 
significance of rejecting the null hypothesis, $r=0$, with the same experimental setup in the left panel. $r^{\rm obs}$ is obtained by fitting the blue points in the left panel to the theory spectrum and the vertical dashed lines are the $2\,\sigma$ bounds.
}
\label{fig:reconst}
\end{figure*}

\begin{figure*}[t]
\centering
\includegraphics[width=88mm]{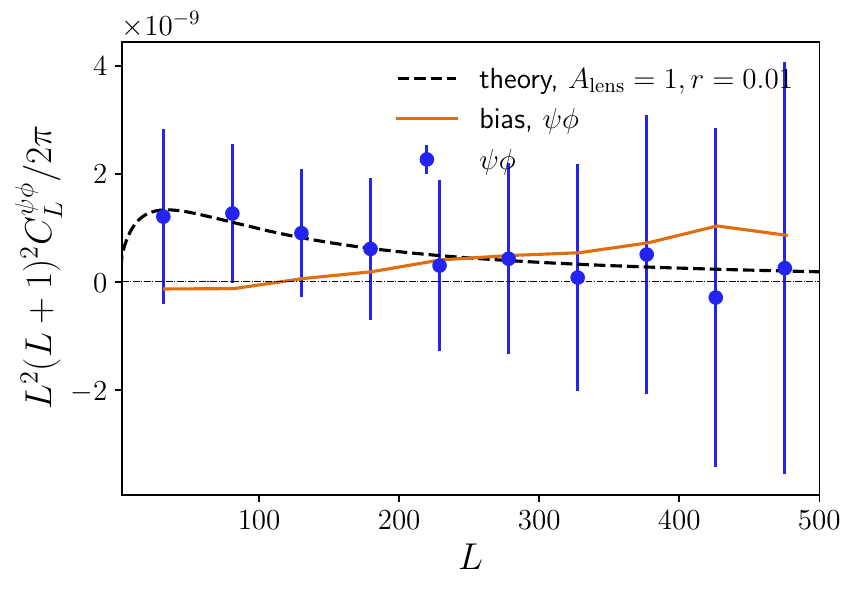}
\includegraphics[width=88mm]{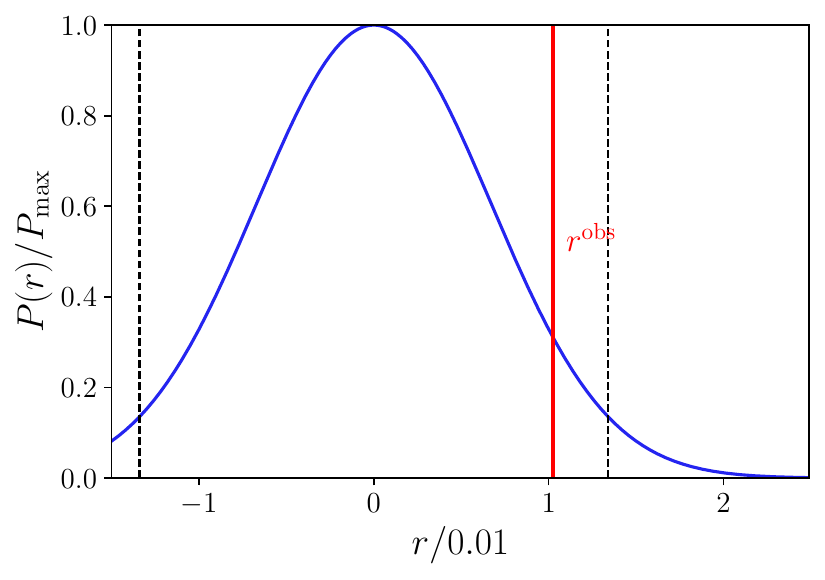}
\caption{
Same as Fig.~\ref{fig:reconst} but for the LiteBIRD case with $r=0.01$.
}
\label{fig:reconst:LBS4W}
\end{figure*}

\begin{figure*}[t]
\centering
\includegraphics[width=88mm]{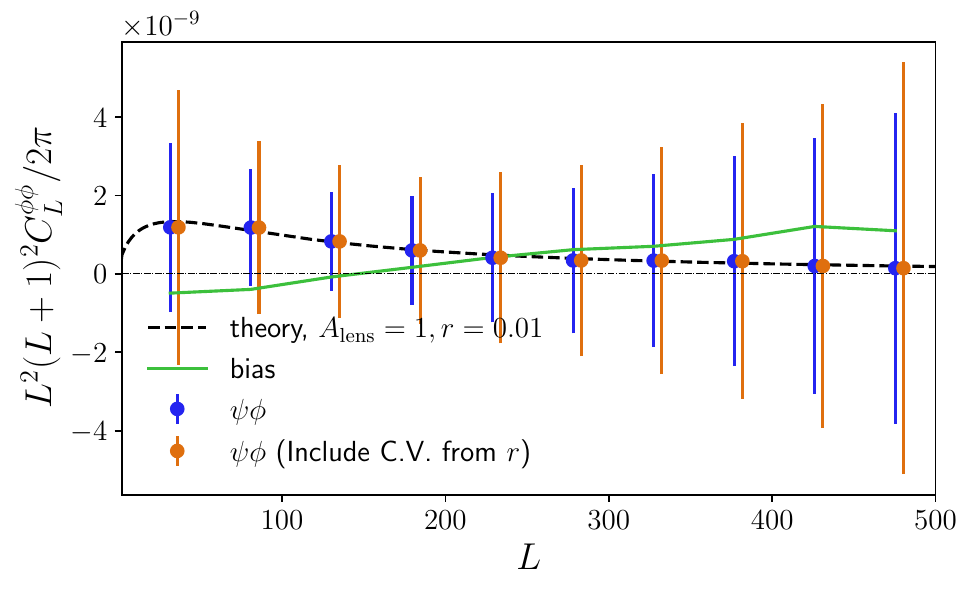}
\caption{
Same as Fig.~\ref{fig:reconst:LBS4W}, but with the $C^{-1}$ filtering including the contributions from the primordial $B$-mode power spectrum with $r=0.01$. The error bars of the blue and orange data points are estimated from simulations with $r=0$ and $r=0.01$, respectively. 
}
\label{fig:reconst:fwr}
\end{figure*}

\begin{figure*}[t]
\includegraphics[width=88mm]{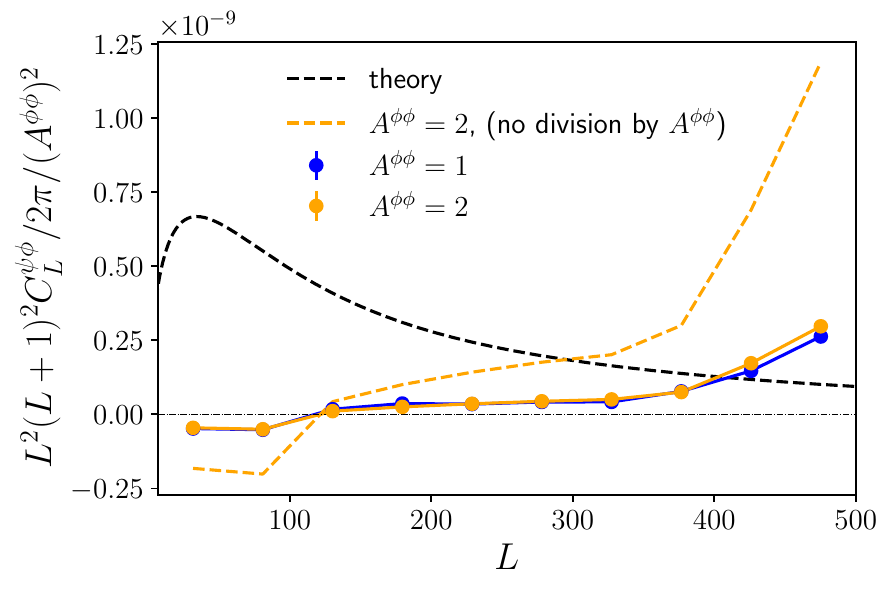}
\includegraphics[width=88mm]{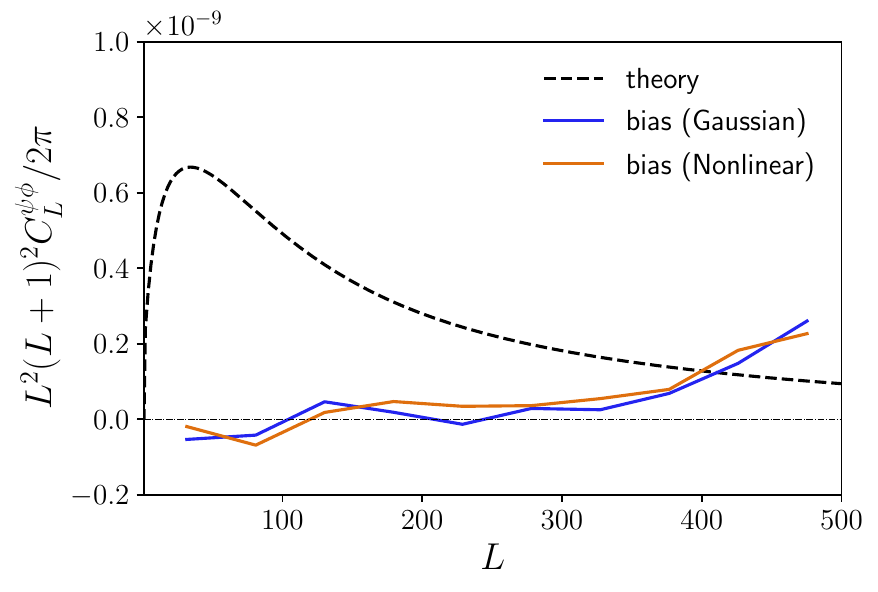}
\caption{
{\it Left}: residual bias terms divided by $(A^{\phi\phi})^2$ where $A^{\phi\phi}$ is the overall amplitude of the lensing potential power spectrum compared to the baseline power spectrum. We assume the same experimental configuration as in Fig.~\ref{fig:reconst}. We compare the bias terms for simulations with different amplitudes of the input lensing potential, $A^{\phi\phi}=1$ and $2$. The black dashed line is the theory signal spectrum with $r=0.005$. The orange dashed line is the same as the orange solid line but without the division by $(A^{\phi\phi})^2$. The points are plotted with the simulation error bars of $400$ realizations.
{\it Right}: residual bias terms obtained from $100$ realizations of the full-sky non-Gaussian simulation of Ref.~\cite{Takahashi:2017} compared with the Gaussian simulation of $\phi$. The error bars denote the $1\,\sigma$ MC simulation error. 
}
\label{fig:bias-terms}
\end{figure*}

Here, we verify, using Monte-Carlo simulations, that any bias arising from non-IGW contributions (such as higher-order effects) is much smaller than the targeted signal. 

We prepare simulations for the no-IGW case ($r=0$) and nonzero values of $r$. We first generate random Gaussian fields for the scalar and tensor CMB anisotropies, noise, and $\phi$. We then remap the CMB anisotropies by $\phi$ with the {\tt LensPix} package \cite{Lewis:2005} for both the scalar and tensor contributions. We assume two experiments for the noise spectrum; an ideal experiment with $0.5\mu$K-arcmin white noise in polarization and $1$ arcmin Gaussian beam, and the LiteBIRD case as described above. For the LiteBIRD case, we use the post-component-separated noise spectrum computed in the previous section. We obtain the observed CMB map as the sum of the lensed scalar-induced CMB, tensor-induced CMB, and noise. For the reconstructed $\phi$, we add random Gaussian noise to the input full-sky $\phi$ instead of directly adopting the maximum-likelihood method \cite{Hirata:2002:mle,Hirata:2003:mle:pol,Carron:2017} 
We assume the use of CMB multipoles between $50$ and $4096$, but the delensing bias does not appear as we generate the reconstruction noise independently \cite{Namikawa:2017:delens,BaleatoLizancos:2020:delens-bias}. We then construct a lensing $B$-mode template by convolving the observed lensed $E$-modes and lensing potential \cite{Seljak:2003pn,Baleato-Lizancos:2020:LT-delens}, and subtract it from the lensing $B$-modes. We input the delensed $B$-modes to reconstruct $\psi$. 

As we described above the reconstructed $\phi$ has a Gaussian random noise whose power spectrum is expected from the iterative/optimal lensing estimator. For the optimal lensing estimate, we expect that any higher-order bias would not be significant and would not significantly increase the bias we showed in our paper. This is because the iterative estimator minimizes the difference between the best estimate and true $\phi$, reducing the residual lensing potential contributions in the reconstruction noise \cite{Darwish:2024yfo}. We will perform a more thorough investigation in future work. 

Figure \ref{fig:reconst} shows the cross-power spectrum between the reconstructed $\psi$ fields and input lensing potential in full sky for the idealistic noise case. The error bars represent the $1\,\sigma$ error for a single full-sky measurement. We obtain the bias term arising from non-IGW contributions from an average over the simulations without IGWs. The blue data points are obtained from the difference between the no-IGW and $r=0.005$ cases, which removes the bias terms. Note that, in the $C^{-1}$ filtering for $B$-modes in the estimator \eqref{Eq:estg} and its normalization \eqref{Eq:Norm:psi}, we ignore the contribution from any nonzero $r$ to the observed $B$-mode power spectrum. In total, we use $1000$ realizations. We remove multipoles below $\ell\leq 6$ where the scatter becomes significant and divide the multipole range from $\ell = 7$ to $500$ into $10$ equally spaced bins. Note that the results are not sensitive to the choice of multipole bins. We find that the bias is not significant compared to the signal power spectrum. Figure~\ref{fig:reconst} also shows the constraint on $r$ from the $r=0$ simulation to see the significance of rejecting the null hypothesis. We estimate $r$ by
\al{
    \hat{r} = \frac{1}{\sum_{b}a_b} \sum_{b} a_b \frac{\hat{C}^{\psi\phi}_b}{C^{\phi\phi}_b}
    \,,
}
where $b$ denotes the index of the multipole bin, $\hat{C}^{\psi\phi}_b$ is the binned cross-power spectrum between the reconstructed $\psi$ and the input $\phi$, and $C^{\phi\phi}_b$ is the fiducial lensing power spectrum. The coefficients $a_b$ are determined from simulation to optimize the estimate $r$ (e.g. Ref.~\cite{BKVIII}): 
\al{
    a_b \equiv \sum_{b'}\bR{Cov}^{-1}_{bb'} 
    \,,
}
where $\bR{Cov}$ is the covariance matrix for $\hat{C}_b^{\psi\phi}/C_b^{\phi\phi}$. 
We find that the $1\,\sigma$ error obtained from the simulation, $\sigma_r=0.00114$, is close to that obtained in the analytic estimate, $\sigma_r=0.00109$, and these values agree within $\sim 4\%$ which is close to the Monte Carlo simulation error. We also check the covariance matrix of the binned $C_L^{\psi\phi}$ and find that the covariance matrix is close to diagonal. 

Figure~\ref{fig:reconst:LBS4W} shows the LiteBIRD case for $r=0.01$. We find that the bias increases compared to the idealistic case as the noise increases and the delensing becomes inefficient. In contrast, the scatter of the simulation average of the power spectrum is reduced. This is because the idealistic case has relatively significant contributions from the cosmic variance arising from nonzero $r$. 

In practice, we can also include the contributions from nonzero $r$ in the $C^{-1}$ filtering for $B$-modes. Figure~\ref{fig:reconst:fwr} shows our results when we include contributions from non-zero $r$ in the $C^{-1}$ filtering. We here assume the LiteBIRD case. Compared with the left panel of Fig.~\ref{fig:reconst:LBS4W}, the statistical errors in $C_L^{\psi\phi}$ are almost unchanged whether or not we include the nonzero $r$ contributions to the $C^{-1}$ filtering. We also show the errors estimated from the $r=0.01$ simulation, i.e., that includes the cosmic variance from $r$. The cosmic variance slightly increases the error bars. We also checked the case for the idealistic simulation, finding that the cosmic variance from $r$ roughly increases the errors by $50\%$ to a factor of $2$. 

Next, we show how the bias depends on the lensing potential. To see this, we change the overall amplitude of the input lensing potential by $\phi\to\sqrt{A^{\phi\phi}}\phi$. We remap the primary CMB fluctuations with the amplified lensing potential and obtain modified lensed $E$ and $B$ modes. To make the lensing reconstruction map, we add the same Gaussian reconstruction noise to $\sqrt{A^{\phi\phi}}\phi$. We then construct the lensing $B$-mode template from the modified lensed $E$-modes and lensing potential but with the same fiducial Wiener filtering and perform delensing to the modified $B$-modes. Using these delensed $B$-modes, we perform the $\psi$-reconstruction with the fiducial filtering function and normalization. We finally correlated the reconstructed $\psi$ map with the input $\phi$ map. 

The left panel of Fig. \ref{fig:bias-terms} shows the results of the $\psi\phi$ power spectrum with the simulations modified in this way. We find that the bias scales approximately with the square of $A^{\phi\phi}$, i.e., the fourth power of $\phi$. This indicates that the dominant term comes from $C_l^{EE}A^{1/2}_{\rm lens}\phi^4$ as expected. 

We additionally check to what extent non-Gaussian contributions to $\phi$ produce a bias in the cross-power spectrum. We use $100$ realizations of the full-sky raytracing simulation produced by Ref.~\cite{Takahashi:2017}. We replace the input Gaussian lensing map with the non-Gaussian lensing map \cite{Takahashi:2017} and perform the same procedure for obtaining $C_L^{\psi\phi}$. 

The right panel of Fig. \ref{fig:bias-terms} shows the results of the cross-power spectrum now including the nonlinearity in the $\phi$ map. The results show that the non-Gaussianity of $\phi$ has negligible impact. Thus, the terms, $\sim N^E\phi^3$ and $\sim C_l^{EE}A_{\rm lens}\phi^3$, are negligible in the bias.

\section{Summary and Discussion} \label{sec:summary}

In the paper, we first updated the forecast presented in NS23 by 1) using the noise spectrum from the HILC component separation and 2) varying the experimental setup. We found that, combining upcoming experiments (LiteBIRD, ASO, and CMB-S4), we achieve $\sigma_r\simeq 0.0035$. 
We then investigated our method with simulations and demonstrated that the higher-order biases are smaller than the signal if $r\agt0.005$ for future CMB experiments; we conclude that they can be easily modeled and subtracted. We also showed that the higher-order biases are well understood, with the dominant term in the higher-order biases arising from $C_l^{EE}A_{\rm lens}^{1/2}\phi^4$, and with non-Gaussianity in the lensing potential negligible. Overall, our conclusion based on extensive simulated tests is that the method is viable.

Although our method yields a constraint on $r$ that is approximately an order of magnitude weaker than that obtained from the standard $B$-mode power spectrum approach, it remains valuable for several reasons. If a nonzero $r$ is detected by SO in the near future, our method is also expected to detect it in forthcoming CMB experiments such as LiteBIRD and CMB-S4, thereby providing an independent confirmation of the detection of IGWs. But even in the absence of a detection by SO, our approach will be useful in future experiments to assess whether a potential nonzero $r$ originates from IGWs or from residual foreground contamination. The complexity of foregrounds and the efficacy of current cleaning techniques remain uncertain. Given that our method responds differently to foregrounds, it can serve as a valuable cross-check, particularly if foreground mitigation proves challenging and significantly degrades the sensitivity of the standard $B$-mode power spectrum analysis.

We have ignored a possible bias from the non-Gaussianity of the Galactic foregrounds. The cross-power spectrum, $C_L^{\psi\phi}$, is a four-point correlation function if $\phi$ is the reconstructed lensing potential from CMB data. Any non-Gaussian sources would yield a bias in the measurement. We will present the impact of the non-Gaussian Galactic foreground in the cross-power spectrum measurement in our future work. 

We have assumed a full-sky idealistic experiment for our simulations. In more realistic data that contain, e.g., a survey mask, non-Gaussian foregrounds, and inhomogeneous noise, the measurement of $C_L^{\psi\phi}$ can be biased, and the statistical errors could be increased. We also ignore extragalactic foregrounds, which are the possible sources of biases (e.g., Refs.~\cite{vanEngelen:2013rla,Sailer:2020,Sailer:2022,BaleatoLizancos:2022:extFGs}). 

While a detailed investigation of the aforementioned caveats is necessary, we have shown that this observable has the potential to be a useful probe of IGWs, which, due to different sensitivity to many potential sources of systematic errors---in particular foregrounds---can be complementary to standard methods for constraining $r$.


\appendix

\section{Relation to the large-scale polarization reconstruction} \label{app:A}

Here, we show that our method is similar to the large-scale $B$-mode reconstruction using small-scale CMB anisotropies and a lensing mass tracer. The large-scale $E$-mode reconstruction was proposed by Ref.~\cite{Meerburg:2017:recon}; in this method, one uses the reconstructed primary $E$-modes to constrain the optical depth. We can also construct the estimator of the primary $B$-modes as follows:
\al{
    \hat{B}^{\rm rec}_{\l m} = N_{\l}^{BB}\sum_{\l'm'LM}\Wjm{\l'}{L}{\l}{m'}{M}{m}W^+_{\l' L\l}\frac{\hat{B}_{\l'm'}^*}{\hC^{BB}_{\l'}}\frac{C_L^{\grad x}\hat{x}_{LM}^*}{\hC_L^{xx}}
    \,,
}
where the normalization is defined as
\al{
    [N_{\l}^{BB}]^{-1} = \frac{1}{2\l+1}\sum_{\l'L}(W^+_{\l'L\l})^2\frac{1}{\hC^{BB}_{\l'}}\rho_L^2C_L^{\phi\phi}
    \,. 
}
We can then compute the cross-correlation between the observed $B$ modes and reconstructed $B$-modes on large angular scales to constrain $r$. This method is, however, not optimal for constraining $r$ because the cross-correlation does not optimally extract the following term: $\hat{B}\hat{B}^{\rm rec} \sim (B\grad)\hat{B}^{\rm rec}$.

\begin{acknowledgments}
TN acknowledges support from JSPS KAKENHI Grant (No. JP20H05859, JP22K03682, and 24KK0248). IAC acknowledges support from Fundación Mauricio y Carlota Botton and the Cambridge International Trust. BDS acknowledges support from the European Research Council (ERC) under the European Union’s Horizon 2020 research and innovation programme (Grant agreement No. 851274). 
Part of this work uses resources of the National Energy Research Scientific Computing Center (NERSC). The Kavli IPMU is supported by World Premier International Research Center Initiative (WPI Initiative), MEXT, Japan. 
\end{acknowledgments}

\section*{Data Availability}

The data that support the findings of this article are not publicly available upon publication because it is not technically feasible and/or the cost of preparing, depositing, and hosting the data would be prohibitive within the terms of this research project. The data are available from the authors upon reasonable request.

\bibliographystyle{apsrev}
\bibliography{cite}

\end{document}